\documentclass[12pt, a4paper]{article}
\usepackage{epsf}
\usepackage{cite}
\usepackage{amsmath,amssymb}
\input{colordvi.tex}
\usepackage[usenames,dvipsnames]{color}
\usepackage{graphicx}
\usepackage{ascmac}
\usepackage[colorlinks=true,linkcolor=red,urlcolor=blue,citecolor=blue]{hyperref}
\usepackage{tikz} 

\usepackage{tikz-feynhand}
\usepackage[subrefformat=parens]{subcaption}
\usepackage{bm}
\graphicspath{{Figure/}}
\usepackage{indentfirst}

\usepackage{bm}
\usepackage{braket}
\usepackage{cleveref}
\usepackage{siunitx}

\begin{document}

\begin{titlepage}

\begin{center}
\hfill TU-1240\\
\hfill KEK-QUP-2024-0020\\
\hfill KEK-TH-2650\\
\hfill KEK-Cosmo-0356\\
\vskip 0.3in

{\Large \bf
Dual Gravitational Wave Signatures \\ of Instant Preheating
}

\vskip .5in

{\large
Wei-Yu Hu$^{(a)}$, Kazunori Nakayama$^{(b,c)}$, Volodymyr Takhistov$^{(c,d,e,f)}$\\and Yong Tang$^{(g,h,i)}$
}

\vskip .25in

$^{(a)}${\em 
School of Physics, Peking University, Beijing 100871, China}

$^{(b)}${\em 
Department of Physics, Tohoku University, Sendai 980-8578, Japan
}

$^{(c)}${\em 
International Center for Quantum-field Measurement Systems for Studies of the Universe and Particles (QUP), KEK, 1-1 Oho, Tsukuba, Ibaraki 305-0801, Japan
}

$^{(d)}${\em 
Theory Center, Institute of Particle and Nuclear Studies (IPNS), High Energy Accelerator Research Organization (KEK), Tsukuba 305-0801, Japan
}

$^{(e)}${\em Graduate University for Advanced Studies (SOKENDAI), \\
1-1 Oho, Tsukuba, Ibaraki 305-0801, Japan}

$^{(f)}${\em 
Kavli Institute for the Physics and Mathematics of the Universe (WPI), UTIAS
The University of Tokyo, Kashiwa, Chiba 277-8583, Japan
} 

$^{(g)}${\em
School of Astronomy and Space Science, University of Chinese Academy of Sciences (UCAS), Beijing 100049, China
}

$^{(h)}${\em
School of Fundamental Physics and Mathematical Sciences,
Hangzhou Institute for Advanced Study, UCAS, Hangzhou 310024, China
}

$^{(i)}${\em
International Center for Theoretical Physics Asia-Pacific, Beijing 100049, China
}

\end{center}
\vskip .3in

\begin{abstract}
In the instant preheating scenario efficient particle production occurs immediately following the period of inflationary expansion in the early Universe. We demonstrate that instant preheating predicts unique gravitational wave (GW) signals arising from two distinct origins. One source is the bremsstrahlung GWs produced through the decay of superheavy particles, an inevitable consequence of instant preheating. The other is GWs generated from the nonlinear dynamics of the inflaton and coupled scalar fields.  Using numerical simulations, we show that the peak of the GW spectrum shifts depending on the coupling constants of the theory. The detection of these dual GW signatures, characteristic of instant preheating, provides novel opportunities for probing the dynamics of the early Universe.
\end{abstract}

\end{titlepage}

\tableofcontents

\section{Introduction}

Gravitational waves (GWs) can be generated by variety of processes in the early Universe. The resulting GW spectra carry rich information about early Universe physics and high-energy phenomena that could be challenging to probe in laboratories otherwise. Notable cosmological GW sources include primordial (inflationary) cosmological quantum fluctuations~\cite{Starobinsky:1979ty}, topological defects~\cite{Vachaspati:1984gt,Vilenkin:2000jqa}, phase transitions~\cite{Kamionkowski:1993fg} as well as processes from preheating stage after inflation~\cite{Khlebnikov:1997di,Easther:2006vd,Easther:2006gt,Dufaux:2007pt} and oscillon formation~\cite{Zhou:2013tsa,Hiramatsu:2020obh,Lozanov:2022yoy,Lozanov:2023aez,Lozanov:2023knf,Lozanov:2023rcd} (see e.g.~\cite{Maggiore:2018sht} for overview). Distinct source can contribute unique signatures to the GW landscape.
 
A multitude of novel possible GW sources associated with the emission of graviton quanta of gravity has been recently identified. Since gravity couples to matter, when the inflaton or heavy particles in general decay perturbatively into light species, graviton production occurs through the bremsstrahlung processes~\cite{Nakayama:2018ptw, Huang:2019lgd, Barman:2023ymn, Bernal:2023wus, Hu:2024awd}. Graviton emission can also efficiently occur from frequent scatterings involving Standard Model (SM) particles in the early Universe's thermal bath~\cite{Ghiglieri:2015nfa, Ghiglieri:2020mhm, Ringwald:2020ist}. Further, during the reheating phase, coherent oscillations of the inflaton lead to gravitational particle production, which can be interpreted as inflaton annihilation into graviton pairs. Similarly, SM particles can also annihilate into graviton pairs~\cite{Ema:2015dka, Ema:2016hlw, Ema:2020ggo, Ghiglieri:2022rfp, Ghiglieri:2024ghm, Choi:2024ilx, Xu:2024fjl}. More so, if the inflaton or other scalar fields couple to the higher (e.g. quadratic) terms of the Riemann tensor, they can decay into graviton pairs with a potentially large branching fraction~\cite{Ema:2021fdz, Mudrunka:2023wxy, Tokareva:2023mrt}.

In this work we explore GW production and establish novel characteristic features from instant preheating~\cite{Felder:1998vq}, where efficient particle production occurs immediately following the period of inflation within just a few oscillations of the inflaton field.
In this scenario, the mass of daughter particles depends on the inflaton field value and can temporarily become superheavy, even approaching the Planck scale. When coupled to lighter particle species, daughter particles can decay into them.
Recently, in Ref~\cite{Hu:2024awd} we found that instant preheating naturally admits highly efficient production of GWs via bremsstrahlung, which is typically thought to be dramatically suppressed. By investigating
minimal realization scenarios of instant preheating with $\mathcal{O}(1)$ couplings between daughter particles and the inflaton, we showed that as daughter particles become superheavy approaching Planck-scale masses their decays can be a remarkably efficient source of high-frequency bremsstrahlung GWs with typical frequencies in the $\sim 10^9-10^{12}$ Hz range. Intriguingly, this suggests that GW observations could probe and provide unique insights about Planck scale physics, challenging to explore. 

Here we significantly expand on the analysis of Ref.~\cite{Hu:2024awd} and explore minimal instant preheating scenarios with different couplings between daughter particles and the inflaton field. In our study we account for GW production from both bremsstrahlung processes as well as broad parametric resonance~\cite{Kofman:1994rk,Kofman:1997yn} during subsequent preheating dynamics~\cite{Khlebnikov:1997di,Easther:2006vd,Easther:2006gt,Dufaux:2007pt}.
We employ lattice simulation for estimating the latter.
Since quantum GW signals from the bremsstrahlung process are an inevitable consequence of the scenario, as we will demonstrate, the generation of these dual GW signals from distinct sources marks a hallmark of instant preheating.

Previously, lattice simulations of the instant preheating scenario have been carried out in Refs.~\cite{Repond:2016sol, Fan:2021otj} and the resulting GW spectrum has been calculated in Ref.~\cite{Mansfield:2023sqp}. 
While Ref.~\cite{Mansfield:2023sqp} considered the tachyonic preheating~\cite{Dufaux:2006ee} for the daughter particle production, we consider a more conventional scenario with the four point coupling like $\sim \frac{1}{2}\lambda^2\phi^2\chi^2$, where $\phi$ is the inflaton and $\chi$ is a daughter particle. 
Importantly, we also point out that the quantum GW signals from the bremsstrahlung processes, which are not incorporated in lattice simulations, are inevitable consequence of the instant preheating scenario.

The paper is organized as follows. In Sec.~\ref{sec:inst} we give a brief review of the instant preheating scenario and consider minimal model realization. We then discuss how a supersymmetric (SUSY) model can readily avoid a constraint on the coupling responsible for the efficiency of instant preheating. We also outline distinct possible regimes of reheating completion.
In Sec.~\ref{sec:gw} we calculate the GW spectrum predicted in the instant preheating scenario. We consider GWs from bremsstrahlung processes and also compute the GWs from preheating dynamics with lattice simulations.
In Sec.~\ref{sec:con} we conclude and discuss implications of our results.

\section{Instant preheating} \label{sec:inst}

We start with a review of instant preheating~\cite{Felder:1998vq}. We first consider a minimal model realization and identify restrictions on the parameters responsible for driving efficiency of preheating. Subsequently, we show how a minimal model based on supersymmetry (SUSY) can alleviate these restrictions and naturally achieve efficient instant preheating.

\subsection{Minimal model realization}
\label{ssec:minmod}

Let us consider a setup described by the Lagrangian
\begin{align} \label{eq:lag}
	\mathcal L =\frac{1}{2}(\partial \phi)^2+\frac{1}{2}(\partial \chi)^2 + i\overline\psi\gamma^\mu \partial_\mu\psi -\frac{1}{2}m_\phi^2\phi^2 - \frac{1}{2}\lambda^2\phi^2\chi^2 + y\chi \bar\psi\psi,
\end{align}
where $\phi$ is the inflaton field with mass $m_{\phi}$, $\chi$ is a scalar field, $\psi$ is a Dirac fermion, $\lambda$ and $y$ are coupling constants. 
Since we are interested in the post-inflationary dynamics, for simplicity we assume that the inflaton potential is well-approximated by a quadratic one.
In the regime of inflation, the scalar potential may be modified in order to be consistent with the cosmic microwave background observation~\cite{Planck:2018jri}, however this is not essential for our discussion.

The interaction coupling $\lambda$ in Eq.~\eqref{eq:lag} induces a radiative correction to the inflaton potential through the Coleman-Weinberg contributions
\begin{align} \label{eq:cw}
	V_{\rm CW}= \frac{m_\chi^4(\phi)}{64\pi^2}\left[ \ln\frac{m_\chi^2(\phi)}{\mu^2}-\frac{3}{2}\right],~~~~~~m_\chi(\phi) = \lambda\phi,
\end{align}
where $\mu$ is the renormalization scale.
Imposing a requirement that large radiative corrections do not spoil the inflaton dynamics translates into an upper bound of $\lambda \lesssim 10^{-3}$.

Typically the inflaton field value is of the order of the Planck scale $M_{\rm Pl} \simeq 2.4\times 10^{18}\,$GeV at the end of slow-roll phase. Thus, the inflaton oscillation amplitude is also typically of the order of $\phi_i \sim M_{\rm Pl}$ in the very first oscillation.
If $\lambda \phi_i \gg m_\phi$, this inflaton oscillation leads to a non-perturbative particle production of $\chi$ with a broad spectrum. The number density of the produced $\chi$ particles in a half-oscillation of the inflaton is estimated as~\cite{Kofman:1997yn}
\begin{align}
	n_\chi \simeq \left(\frac{k_*}{2\pi}\right)^3,~~~~~~k_* = \sqrt{\lambda m_\phi \phi_i},
\end{align}
where $k_*$ represents a typical wavenumber of the produced $\chi$ particles. 
It means that the typical wavenumber of $\chi$ can be much larger than the inflaton mass $m_\phi$.
In the absence of the Yukawa coupling $y$, $\chi$ particles are stable and the particle production in the next half-oscillation of the inflation field receives a Bose enhancement effect. This leads to the parametric resonance that can be broad for the $\chi$ particle~\cite{Kofman:1994rk,Kofman:1997yn}. 

In the instant preheating scenario~\cite{Felder:1998vq}, the $\chi$ field has a sizable Yukawa coupling to lighter fermions and hence it can (perturbatively) decay into fermion pairs.
In the most efficient realization, it is even possible that the produced $\chi$ particles during the fist half-oscillation of the inflaton can completely decay before the next half-oscillation.
Hence, in this scenario some fraction of the inflaton energy rapidly converts into radiation and preheating can be nearly instantaneous~\cite{Felder:1998vq}.
The perturbative decay rate of $\chi$ is given by
\begin{align}
	\Gamma_{\chi\to\bar\psi\psi} = \frac{y^2 \lambda \phi(t)}{8\pi} \simeq \dfrac{y^2 \lambda m_{\phi}\phi_i t}{8 \pi}~,
	\label{Gamma}
\end{align}
where $t$ is time.
Requiring that the decay occurs within a single oscillation period $1/m_{\phi}$ of the inflaton imposes a constraint of
\begin{align}
	m_\phi \lesssim \frac{ y^2 \lambda \phi_i}{8\pi}.  \label{y_constraint}
\end{align}
For the canonical values of $m_\phi \simeq 10^{13}\,$GeV and $\phi_i\simeq M_{\rm Pl}$ Eq.~\eqref{y_constraint} translates into $y^2 \lambda \gtrsim 10^{-4}$.

The mass of $\chi$ at the instance of decay, $t=t_{\rm dec}$, can be estimated as
\begin{align}
	m_\chi (t_{\rm dec}) \simeq \sqrt{\frac{8\pi \lambda m_\phi \phi_i}{y^2}} \lesssim \lambda \phi_i,~~~~~~t_{\rm dec} \simeq \sqrt{\frac{8\pi}{y^2\lambda m_\phi \phi_i}}
	\label{mchi_dec}
\end{align}
Thus, $m_\chi$ can become superheavy approaching Planck scale when it decays if $\lambda\sim \mathcal O(1)$. As we demonstrate below sizable $\lambda$ couplings can be naturally realized in the context of SUSY models without restrictions of Eq.~\eqref{eq:cw}\footnote{
    Sizable $\lambda$ couplings can also be realized by introducing a large nonminimal coupling of $\phi$ to gravity like in Higgs inflation~\cite{Bezrukov:2007ep} or singular kinetic term like in attractor models~\cite{Kallosh:2013yoa}. Instant preheating in the context of Higgs inflation has been considered in Refs.~\cite{Bezrukov:2008ut,Garcia-Bellido:2008ycs}, however potential theoretical problems with such realizations have been identified~\cite{Ema:2016dny}.
}.
When the upper bound of Eq.~\eqref{y_constraint} is saturated, the inflaton energy loss is most efficient. 
The energy loss of the inflaton in one half-oscillation can then be estimated as
\begin{align}
	\frac{\delta\rho_\phi}{\rho_\phi} \simeq \frac{\rho_\chi(t_{\rm dec})}{\rho_\phi} \sim \frac{m_\chi(t_{\rm dec}) n_\chi}{\rho_\phi}
	\sim \frac{\lambda^2}{4\pi^3}\sqrt{\frac{8\pi}{y^2}} \lesssim \mathcal O(1)\times \lambda^{5/2}.
	\label{rhochi_rhophi}
\end{align}
In the last inequality we used Eq.~\eqref{y_constraint} and considered $m_\phi \simeq 10^{13}\,{\rm GeV}$, $\phi_i \simeq M_{\rm Pl}$. 

The whole cascade process can be viewed as a two-step decay of the inflaton $\phi\to\chi\to\psi$ and we can define an effective inflaton decay rate. This can be estimated (see also~Ref.~\cite{Felder:1998vq}) as
\begin{align}
	\Gamma_\phi = \frac{\delta\rho_\phi}{\rho_\phi}\frac{m_\phi}{\pi} \simeq \frac{\lambda^2 m_\phi}{4\pi^4}\sqrt{\frac{8\pi}{y^2}},
	\label{Gamma_phi}
\end{align}
where $\delta\rho_\phi$ is the inflaton energy transferred to $\chi$ in one oscillation.
Taking into account the restriction on $y$ from Eq.~\eqref{y_constraint},~we find~\cite{Hu:2024awd} that the maximum efficiency for the $\chi$ decay and resulting bremsstrahlung GWs is achievable for $\lambda \sim \mathcal O(1)$ and $y \sim \mathcal O(10^{-2})$.

\subsection{Supersymmetric model}

Let us now discuss a realization of instant preheating in a minimal SUSY model. 
We consider the Kahler potential $K$ and superpotential $W$ given by
\begin{align}
	& K = \frac{1}{2}(\phi+\phi^\dagger)^2 + |X|^2 + |\chi|^2 + |\Psi|^2, \label{K}\\
	& W = m_\phi X \phi + \lambda \phi \chi^2 + y \chi \Psi^2,  \label{W}
\end{align}
where $\phi$ is an inflaton chiral superfield, $X$ is a so-called stabilizer chiral superfield, $\chi$ and $\Psi$ are also chiral superfields taking the roles of $\chi$ and $\psi$ discussed in the non-SUSY minimal model of Sec.~\ref{ssec:minmod}.
We take $\lambda$ and $y$ couplings as real and positive without loss of generality.
Each chiral superfield contains both scalar and fermionic components\footnote{Here, we represent a chiral superfield and its scalar component with same notation.}.
Due to an approximate shift symmetry $\phi\to \phi + iC$ with arbitrary real constant $C$, large field inflation can be realized within the framework of supergravity~\cite{Kawasaki:2000yn}.
As concretely studied in Refs.~\cite{Nakayama:2013jka,Nakayama:2013txa}, the inflaton potential with slight modifications
is consistent with measurements from Planck~\cite{Planck:2018jri}. However, this is not essential in the following discussion. 
We note that the sneutrino chaotic inflation model~\cite{Murayama:2014saa,Evans:2015mta,Nakayama:2016gvg,Kallosh:2016sej} well fits into this framework.

The scalar potential obtained from Eq.~\eqref{K} and Eq.~\eqref{W} is
\begin{align}
	V = m_\phi^2|\phi|^2 + |m_\phi X + \lambda \chi^2|^2 + |\lambda \phi\chi + y\Psi^2|^2 + y^2|\chi|^2|\Psi|^2.
\end{align}
Considering relevant terms around the vacuum $X=\chi=\Psi=0$, we obtain the interaction Lagrangian 
\begin{align}
	  \mathcal L_{\rm int} ~&=  \left[\lambda \phi \tilde\chi \tilde\chi + y(\chi \tilde\psi\tilde\psi + 2 \psi \tilde\chi\tilde\psi) + {\rm h.c.} \right]- V,\\
	  V ~&\simeq m_\phi^2|\phi|^2 + \lambda^2 |\phi|^2|\chi|^2 + \lambda y (\phi \chi \Psi^{\dagger 2} + {\rm h.c.}),
\end{align}
where tilde indicates the fermionic component of each chiral superfield, which represents a Weyl fermion in a two-component notation.
Thus, in our SUSY model realization of instant preheating both the scalar and fermionic components of $\chi$ are produced by the first half-oscillation of the inflaton with similar amount\footnote{
    Small Yukawa coupling $y$ can result in parametric resonance, which will Bose-enhance production of the scalar component while the fermionic component is saturated due to Pauli blocking.	
}.
The scalar component decays through $\chi \to \Psi\Psi$ or $\chi\to\tilde\psi\tilde\psi$ while the fermionic component through $\tilde\chi\to \psi\tilde\psi$.
The processes result in particle production and decays 
with same orders of magnitude in efficiency.

As discussed, the estimates of the non-SUSY scenario in Sec.~\ref{ssec:minmod} are also applicable to the SUSY scenario up to additional corrections in numerical factors. However, in our SUSY realization large coupling constant $\lambda$ does not spoil the inflaton dynamics since radiative corrections from a scalar $\chi$ and fermion $\tilde\chi$ cancel each other\footnote{More precisely, there is a finite radiative correction to the inflaton potential due to SUSY breaking effects, which we do not specify. However, this is negligible considering that the SUSY breaking scale is significantly below the inflaton mass scale.
}.
Hence, we can readily consider $\lambda \sim \mathcal O(1)$, neglecting Coleman-Weinberg contributions of Eq.~\eqref{eq:cw}.

Let us briefly comment on the cosmological problem of gravitino overproduction~\cite{Khlopov:1984pf,Ellis:1984eq}.
Efficient instant preheating implies high reheating temperatures after inflation, which could result in efficient production of gravitino superpartners of gravitons.
If the gravitino mass is around TeV-scale, the decays of produced gravitinos around the Big-Bang nucleosynthesis (BBN) epoch can significantly affect the light element abundances~\cite{Kawasaki:2004qu,Jedamzik:2006xz,Kawasaki:2017bqm}.
To avoid this, we consider that gravitinos are sufficiently heavy (e.g. with mass of 100\,TeV or above) so that they decay well before the onset of BBN\footnote{
Still, the lightest SUSY particle (LSP) may be overproduced by the late time gravitino decays, which may exceed the observed dark matter abundance. To avoid the LSP overproduction, one can consider a small R-parity breaking resulting in unstable LSP.
}. We leave detailed analysis of such considerations for future work.

\subsection{Completion of reheating}

We now summarize how reheating completes in our scenario in the early Universe after inflation.
If instant preheating is sufficiently efficient, the radiation-dominated phase begins soon after inflation.
If instant preheating is not efficient, it is possible that it produces only a fraction of the total radiation energy density and eventually parametric resonance leading to $\chi$ particle production begins. In this case, the radiation-dominated Universe occurs after the parametric resonance. 

We show how key parameters of our scenario determine the reheating outcome.
First, we consider condition for instant preheating to occur just after inflation. This is given by Eq.~\eqref{y_constraint}, which we can rewrite as
\begin{align}
	y \gtrsim 10^{-2} \lambda^{-1/2}
\end{align}
for characteristic values of $m_\phi \simeq 10^{13}$\,GeV and $\phi_i \simeq M_{\rm Pl}$, as before. If this condition is not satisfied, the produced $\chi$ particles do not decay within one inflaton oscillation and parametric resonance resulting in $\chi$ particle production happens. Hence, there would be no instant preheating epoch.

Second, we consider condition for instant preheating to be sufficiently efficient such that almost all of the inflaton energy is converted into radiation within just a few inflaton oscillations. This requires $\Gamma_\phi \gtrsim m_\phi$, where $\Gamma_\phi$ is given by Eq.~\eqref{Gamma_phi}, which is rewritten as
\begin{align}
	y \gtrsim 10^{-2} \lambda^2.
\end{align}
If this is not satisfied, the instant preheating regime lasts long time. Eventually the reheating is completed either during this regime or after the parametric resonance sets in.

Third, let us consider the condition for parametric resonance to turn on. 
Here, as we discuss in more detail in Sec.~\ref{ssec:paramres}, oscillations of inflaton field around potential minimum can lead to a exponentially-growing instability modes within a resonance band, described by exponential growth factor $\mu_k$. Hence, parametric resonance works efficiently when it overcomes decays described by $\Gamma_{\rm dec} = 1/t_{\rm dec}$ and Hubble expansion, $\mu_k m_{\phi} \gtrsim 3 H + \Gamma_{\rm dec}$. For broad resonance, which will be our focus, we can crudely approximate the regime by considering $\mu_k$ not far from unity~\cite{Kofman:1994rk,Kofman:1997yn} and that relevant Hubble parameter is almost always subdominant to $m_{\phi}$, obtaining simplified condition $m_{\phi} t_{\rm dec} \gtrsim 1$.

Denoting the inflaton amplitude by $\widetilde\phi \simeq \phi_i (m_\phi t)^{-1}$, we derive the cosmic time at which the parametric resonance begins $(t_{\rm P.R.})$ from the condition $m_\phi t_{\rm dec} \gtrsim 1$, where $t_{\rm dec}$ is given by Eq.~\eqref{y_constraint} with the replacement $\phi_i \to \widetilde \phi$. Then, we find
\begin{align}
	t_{\rm P.R.} \simeq \frac{1}{8\pi} \frac{y^2 \lambda \phi_i}{m_\phi^2}.
\end{align}
On the other hand, the cosmic time when almost all the inflaton energy is converted into radiation due to the instant preheating effect ($t_{\rm I.R.}$) is estimated by $\Gamma_\phi \gtrsim H$, where $\Gamma_\phi$ is given by Eq.~(\ref{Gamma_phi}) with replacing $\phi_i \to \widetilde\phi$. From this we obtain
\begin{align}
	t_{\rm I.R.} \simeq \frac{\pi^4}{\sqrt{3\pi}} \frac{y \phi_i}{\lambda^2 m_\phi M_{\rm Pl}}.
\end{align}
If $t_{\rm I.R.} < t_{\rm P.R.}$, the reheating is completed due to the instant preheating effect. This condition is rewritten as
\begin{align}
	y \gtrsim 10^{-2} \lambda^{-3}.
\end{align}
On the other hand, if $t_{\rm I.R.} > t_{\rm P.R.}$, the parametric resonance is turned on before the instant preheating effect completes the reheating.
The evolution after the parametric resonance is highly nonlinear and complicated, and also depends on the properties of $\psi$ particle.
In this study we simply assume that the Universe becomes effectively radiation-dominated after the parametric resonance is turned on.

\begin{figure}[t]
    \centering
    \includegraphics[width=1\textwidth]{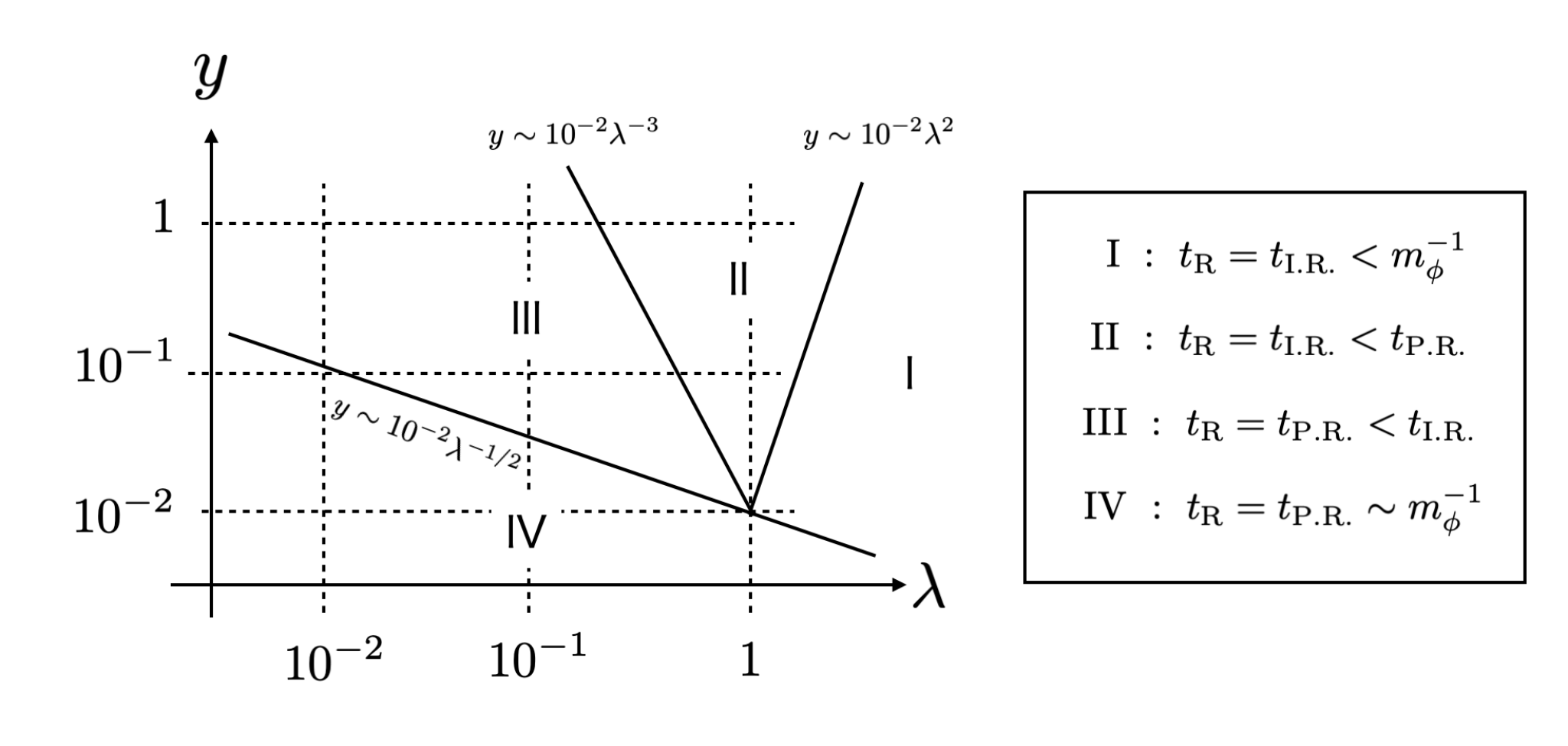}
  \caption{Parameter dependence of the early Universe reheating scenario in $(\lambda,y)$ plane. In region I, instant preheating is strong enough and reheating is completed within a few oscillations after inflation. In region II, the instant preheating not strong enough but still the reheating is completed before the parametric resonance is turned on. In region III, the parametric resonance is turned on after the epoch of instant preheating. In region IV, the instant preheating does not happen at all and parametric resonance happens just after inflation. 
  Characteristic values $m_\phi \simeq 10^{13}$\,GeV and $\phi_i \simeq M_{\rm Pl}$ are assumed. } 
  \label{fig:param}
\end{figure}

In Fig.~\ref{fig:param} we summarize distinct regimes of reheating and their estimated dependence on the values of parameters $\lambda$ and $y$. In region I, the instant preheating is strong enough and the reheating is completed within a few oscillations after inflation. In region II, the instant preheating is not strong enough but still the reheating is completed before the parametric resonance is turned on. In region III, the parametric resonance is turned on after the epoch of instant preheating. In region IV, the instant preheating does not happen at all and parametric resonance happens just after inflation. 
Characteristic values of $m_\phi \simeq 10^{13}$\,GeV and $\phi_i \simeq M_{\rm Pl}$ are assumed.
We stress that the boundaries of these regions are approximate and the distinction among them may not be very strict.
The time at which the radiation-domination epoch occurs ($t_{\rm R}$) can be estimated as
\begin{align}
	t_{\rm R} \simeq \begin{cases}
		m_\phi^{-1} ~~~&{\rm region~I~or~IV} \\
		t_{\rm I.R.}~~~&{\rm region~II} \\
		t_{\rm P.R.}~~~&{\rm region~III} 
	\end{cases}.
\end{align}
We are interested in a scenario with instant preheating, namely regions I, II or III.
Below, we proceed to estimate the GW spectrum expected in these regimes. 

\section{Gravitational waves from instant preheating} \label{sec:gw}

\subsection{Graviton bremsstrahlung processes}

As explained in Sec.~\ref{sec:inst}, during the inflaton oscillations just after inflation, $\chi$ particles are efficiently produced and decay into $\psi$ particles.
Let us consider the graviton bremsstrahlung process as $\chi\to \psi + \bar\psi + h$ with $h$ being the graviton. The differential decay rate of this three body decay process is given by~\cite{Nakayama:2018ptw,Huang:2019lgd,Barman:2023ymn,Bernal:2023wus}
\begin{align}
	E\frac{d\Gamma_{\chi\to\psi\psi h}}{dE} =&~\frac{y^2}{64\pi^3}\frac{m_\chi^3}{M_{\rm Pl}^2}  
	\left[
		(1-2x)\left(8xw^2+2x(x-1)-8w^4-2w^2+1\right)\alpha \right.\nonumber\\
		&~+ 4w^2\left((5-8x)w^2-(x-1)^2-4w^2\right) \ln \frac{1+\alpha}{1-\alpha} 
	],
\end{align}
where $E$ is the graviton energy, $x\equiv E/m_\chi$, $w\equiv m_\psi/m_\chi$ and $\alpha\equiv \sqrt{1-4w^2/(1-2x)}$. Taking $w=0$ ($m_\psi=0$), this becomes
\begin{align}
	E\frac{d\Gamma_{\chi\to\psi\psi h}}{dE} =\frac{y^2}{64\pi^3}\frac{m_\chi^3}{M_{\rm Pl}^2}(1-2x)(1-2x+2x^2).
\end{align}
Thus, the branching ratio of $\chi$ to the graviton bremsstrahlung process is
\begin{align} \label{eq:gravbrem}
	{\rm Br}_{\chi\to\psi\psi h}\simeq \frac{1}{8\pi^2} \left(\frac{m_\chi}{M_{\rm Pl}}\right)^2.
\end{align}
We highlight that Eq.~\eqref{eq:gravbrem} depends only on $m_\chi$, but not on the Yukawa coupling $y$. Therefore, the abundance of bremsstrahlung GWs become larger for heavier $\chi$.
Thus, the instant preheating scenario provides excellent opportunity to have maximally enhanced GW production.
By using Eq.~\eqref{mchi_dec} and noting that the inflaton amplitude decreases as $\sim \phi_i (m_\phi t)^{-1}$ for cosmic time $t$, the branching ratio is estimated as
\begin{align}
	{\rm Br}_{\chi\to\psi\psi h}\simeq \frac{\lambda}{\pi y^2}\frac{m_\phi \phi_i}{M_{\rm Pl}^2}\frac{1}{m_\phi t}.
\end{align}

We can then estimate the present day energy spectrum of the bremsstrahlung graviton emission as
\begin{align}
	\frac{d\rho_{\rm GW}}{d\ln E_0}
	= E_0\int dt \, n_\chi(t) \left(\frac{a(t)}{a_0}\right)^3 \frac{d\Gamma_{\chi\to\psi\psi h}(t)}{d\ln E}
	= E_0\int \frac{dz}{H(z)} \, \frac{n_\chi(z)}{(1+z)^4}\frac{d\Gamma_{\chi\to\psi\psi h}(z)}{d\ln E},
	\label{rhoGW}
\end{align}
where $\rho_{\rm GW}$ is the GW energy density, $E_0$ is the present energy, $n_\chi$ is the number density of $\chi$ and $E(t)=E_0 a_0/a(t)$. The subscript 0 indicates the value at present, redshift $z=0$.
As shown below, the dominant contribution comes from decays around the completion of reheating, $t=t_{\rm R}$ (or correspondingly $z=z_{\rm R}$), at all frequencies.
Then, the GW spectrum is given by
\begin{align}
	\frac{d\Omega_{\rm GW}}{d\ln E_0} 
	=\Omega_{\rm rad} \left(\frac{g_*}{g_{*0}}\right)\left(\frac{g_{*s0}}{g_{*s}}\right)^{\frac{4}{3}} 
	\left(\frac{\rho_\chi}{\rho_{\rm tot}}\right)_{z_{\rm R}} {\rm Br}_{\chi\to\psi\psi h}
	\times x (1-2x)(1-2x+2x^2).
	\label{Ogw}
\end{align}
where $x= E_0/E_{\rm peak}$, $E_{\rm peak}\equiv m_\chi/(1+z_{\rm R})$, $\rho_{\rm tot}$ denotes the total energy density at the  redshift $z=z_{\text{R}}$, $\Omega_{\rm rad}\sim 8\times 10^{-5}$, $g_*$ and $g_{*s}$ represent the effective relativistic degrees of freedom for energy and entropy density, respectively.  
The energy fraction of $\chi$ is estimated from Eq.~\eqref{Gamma_phi} as
\begin{align}
    \left(\frac{\rho_\chi}{\rho_{\rm tot}}\right)_{z_{\rm R}} \simeq (m_\phi t_{\rm R}) \times \sqrt{\frac{1}{2\pi^5}}\frac{\lambda^2}{y},
    \label{fracrhochi}
\end{align}
where the factor $m_\phi t_{\rm R}$ accounts for the fact that the energy density of $\rho_\chi$ is produced, and subsequently decays, around $m_\phi t_{\rm R}$ times within one Hubble time when $t=t_{\rm R}$.

The peak value of the spectrum can be approximatly estimated as 
\begin{equation}
\Omega_{\rm GW, peak} \sim \Omega_{\rm rad}\times {\rm Br}_{\chi\to\psi\psi h} \times \Big(\dfrac{\rho_\chi}{\rho_{\rm tot}}\Big)_{z_{\rm R}}    
\end{equation}
and hence it is proportional to $\Omega_{\rm GW, peak}\propto (\lambda/y)^3$. The typical peak GW spectrum frequency can be found to be
\begin{align}
	f_{\rm peak} = \frac{E_{\rm peak}}{2\pi} \simeq 4.6\times 10^{13}\,{\rm Hz} \left(\frac{m_\chi(t_{\rm R})}{M_{\rm Pl}}\right) \left(\frac{10^{15}\,{\rm GeV}}{T_{\rm reh}}\right),
\end{align}
where $T_{\rm reh}$ is the reheating temperature at $t=t_{\rm R}$, below which the Universe is radiation-dominated. Here,   $m_\chi(t_{\rm R})$ is given by
\begin{align} \label{eq:mchireh}
	m_\chi (t_{\rm R}) \simeq \sqrt{\frac{8\pi\lambda\phi_i}{y^2 t_{\rm R}}}.
\end{align}

\begin{figure}[tbp]
\centering
	\includegraphics[width=0.48\textwidth, height=0.35\textwidth]{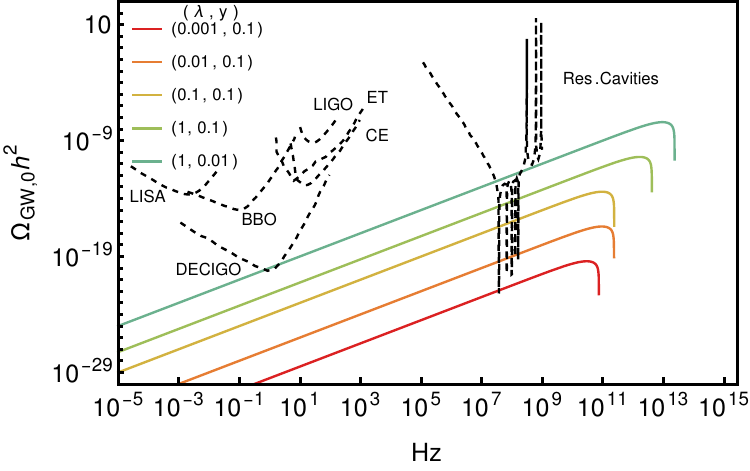}  
 \hspace{1em}
    \includegraphics[width=0.48\textwidth, height=0.35\textwidth]{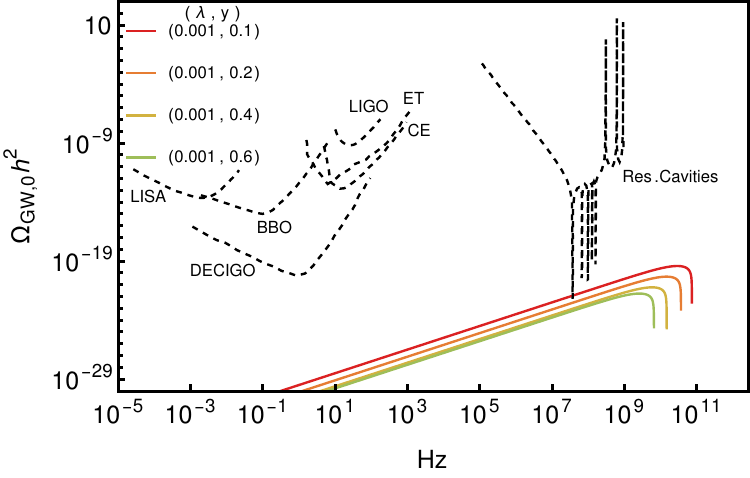}  
\caption{Expected GW energy spectrum today, stemming from graviton bremsstrahlung (solid colored lines) associated with decay of heavy particles for various values of $(\lambda, y)$. Sensitivity curves (dashed lines) of proposed and existing GW detectors, including the fifth observing run (O5) of the aLIGO-Virgo detector network~\cite{LIGOScientific:2016fpe}, Laser Interferometer Space Antenna (LISA)~\cite{LISA:2017pwj}, Cosmic Explorer (CE)~\cite{Reitze:2019iox}, Einstein Telescope (ET)~\cite{Punturo:2010zz}, Big Bang Observer (BBO)~\cite{Harry:2006fi}, DECi-hertz Interferometer Gravitational wave Observatory (DECIGO)~\cite{Seto:2001qf} and resonance cavities~\cite{Herman:2020wao,Herman:2022fau} are also displayed.}
  \label{fig:GW_brems}
\end{figure}

The left panel of Fig.~\ref{fig:GW_brems} displays GW energy spectra from graviton bremsstrahlung emission in the instant preheating scenario for various parameter combinations $(\lambda, y)$ along with sensitivities of proposed and existing GW detectors. Here we have assumed characteristic values $m_\phi=10^{13}\,{\rm GeV}$, $\phi_i=M_{\rm Pl}$. We observe that the peak position of the GW energy spectrum is determined by $m_{\chi}$ and generally lies beyond $10^{10}$\,Hz. As $\lambda$ increases, the $\chi$ particles become heavier, which in turn raises the peak frequency of the GW spectrum. Furthermore, an increases in $\lambda$ leads to a higher transfer of energy density from the inflaton field to $\chi$ field during instant preheating. This results in a greater number of $\chi$ particles, thereby increasing the peak amplitude of the GW spectrum correspondingly. Intriguingly, we note that recently proposed GW detectors based on resonant cavities~\cite{Herman:2020wao,Herman:2022fau} can be sensitive to GWs from graviton bremsstrahlung associated with heavy particle decays during instant preheating, if such experimental setups can be realized. Additionally, at lower frequencies, instant preheating can result in potentially observable GW signatures in experiments such as DECIGO~\cite{Seto:2001qf}.

The right panel of Fig.~\ref{fig:GW_brems} shows the GW energy spectra for various parameter $y$ with $\lambda=10^{-3}$. We can see that as the parameter $y$ increases, as described by Eq.~\eqref{mchi_dec} and Eq.~\eqref{eq:mchireh}, the $\chi$ particles will decay earlier and the resulting mass for the $\chi$ particles is lighter. Consequently, both the peak frequency and the amplitude of the GW spectrum will decrease. As we demonstrate, although the mass of $\chi$ particles does not reach the Planck scale, they still generate a significant GW signal.  

\subsubsection{Gravitational wave dominance at reheating epoch}

We now demonstrate that the GW spectrum of Eq.~\eqref{rhoGW} is dominated by production around $t\simeq t_{\rm R}$ at all frequencies\footnote{
    Strictly speaking, in the parameter region III of Fig.~\ref{fig:param}, bremsstrahlung GWs are also produced even for $t>t_{\rm R}$ after the parametric resonance is turned on. For simplicity, throughout this work, we only consider direct production of GWs from nonlinear $\chi$ fluctuations during the parametric resonance regime. 
}.
We first note that the bremsstrahlung GW spectrum produced at any time $t$ is proportional to $E_0$ for $E_0 \lesssim E_{\rm peak}(t)$ and falls off rapidly around $E \sim E_{\rm peak}(t)$.
Let us denote the peak value of the GW spectrum produced at each cosmic time $t$ by $\Omega_{\rm GW, peak} (t)$.
If the peak energy becomes larger with increased time $t$, it is clear that the contributions from later time dominate at all frequencies, considering that power exponent $n>1$ where $n$ is defined as $\Omega_{\rm GW, peak} (t) \propto \left(E_{\rm peak}(t)\right)^n$. 

The peak energy of the GW spectrum produced at the cosmic time $t$ is given by
\begin{align}
	E_{\rm peak}(t) = m_\chi(t) \frac{a(t)}{a_0} \propto t^{1/6},
\end{align}
where we have used $m_\chi(t) \simeq \sqrt{8\pi\lambda\phi_i/y^2 t}$.
The GW energy spectrum from the cosmic time $t$, within one Hubble time, can be estimated as
\begin{align}
	\frac{d\Omega_{\rm GW}}{d\ln E_0} (t) \propto \frac{E_0}{E_{\rm peak}}\times {\rm Br}_{\chi\to\psi\psi h} \times \left(\frac{\rho_\chi(t)}{\rho_{\rm tot}(t)}\right)\times
	\left(\frac{a(t)}{a(t_{\rm R})}\right) \propto t^{2/3}
\end{align}
for $E_0 = E_{\rm peak}$.
To obtain the right hand side of equation above, we can estimate each contributing term as follows. By noting that $ {\rm Br}_{\chi\to\psi\psi h} \propto \left(m_\chi(t)/M_{\rm Pl}\right)^2$, we find $ {\rm Br}_{\chi\to\psi\psi h} \propto t^{-1}$. Since the $m_\phi t$ factor contributes as in Eq.~\eqref{fracrhochi}, we find $\rho_\chi(t)/\rho_{\rm tot}(t) \sim t$. Finally, $a(t)/a(t_{\rm R}) \propto t^{2/3}$.
Therefore, we find 
\begin{equation}
    \Omega_{\rm GW, peak} (t) \propto \left(E_{\rm peak}(t)\right)^4~,
\end{equation}
completing the proof.

\subsection{Parametric resonance}
\label{ssec:paramres}

In the instant preheating scenario, besides GWs generated from graviton bremsstrahlung associated with particle decays during the first oscillation of the inflaton field, there is an additional source of GWs generated through nonlinear dynamics of scalar fields in the later stages of preheating~\cite{Khlebnikov:1997di,Easther:2006vd,Easther:2006gt,Dufaux:2007pt}. 
Note that even if there is no parametric resonant enhancement of $\chi$ fluctuation at earlier stages during first few inflaton oscillations due to $\chi$ decays into $\psi$, at later preheating stages the inflaton amplitude decreases and hence the $\chi$ decay rate also decreases such that the conditions for parametric resonance can be satisfied.
Here, we employ numerical lattice methods to calculate the GW signals 
generated through parametric resonance in the context of instant preheating scenario. 

Incorporating the effects of $\chi$ decays into $\psi$ within lattice simulations is technically challenging. Instead, we account for these effects by introducing an effective decay term in the equation of motion of $\chi$, as was done in previous studies~\cite{Repond:2016sol, Fan:2021otj}. In Ref.~\cite{Mansfield:2023sqp} the resulting GW spectrum has been numerically analyzed employing similar strategy in the context of tachyonic preheating for $\chi$ production. Instead, we consider a more conventional scenario with the four-point interaction coupling $\sim \frac{1}{2}\lambda^2\phi^2\chi^2$. 

\subsubsection{Conditions for resonance}

Firstly, we consider the preheating without the particle decay process. 
The interaction of the inflaton field $\phi$ and the daughter field $\chi$ is given by $\frac{\lambda^2}{2}\phi^2\chi^2$. 
The classical equation of motion (EoM) for the inflaton field $\phi$ and the daughter field $\chi$ are given by
\begin{align}
  \Ddot{\phi}+3H\Dot{\phi}-\frac{1}{a^2}\nabla^2\phi+m_{\phi}^2\phi+\lambda^2\chi^2\phi~&=0, \\
  \label{eq:chi1}
  \Ddot{\chi}+3H\Dot{\chi}-\frac{1}{a^2}\nabla^2\chi+\lambda^2\phi^2\chi~&=0.
\end{align}
where ``.'' denotes derivative respect to the cosmic time $t$, $a$ is the scale factor, $H=\dot{a}/a$ denotes the Hubble rate. 
At the end of the inflation, the inflaton field $\phi$ undergoes oscillations around the minimum of the potential $V\left(\phi\right)$ with a frequency $m_{\phi}$. 
Consequently, the solution for $\phi\left(t\right)$ can be approximately written as,
  \begin{equation}
    \phi\left(t\right)=\Phi\left(t\right)\sin (m_{\phi}t)\, ,
  \end{equation}
  where $\Phi\left(t\right)$ represents oscillation amplitude of $\phi$, which is gradually decaying due to the Hubble friction.
  The equations for the perturbations are
  \begin{align}
      \label{eq:pertchi}  \delta\Ddot{\phi}_{\bm{k}}+3H\delta\Dot{\phi}_{\bm{k}}+\left(\frac{k^2}{a^2}+m_{\phi}^2\right)\delta\phi_{\bm{k}}~&=0, \\
      \delta\Ddot{\chi}_{\bm{k}}+3H\delta\Dot{\chi}_{\bm{k}}+\left(\frac{k^2}{a^2}+\lambda^2\phi^2\left(t\right)\right)\delta\chi_{\bm{k}}~&=0.
  \end{align}

  The EoM for the perturbation $\delta\phi_{\bm{k}}$ indicates a simple oscillating solution with a frequency $\omega_{\bm{k}}=\sqrt{k^2/a^2+m_{\phi}^2}$ damped by the Hubble friction when we neglect the backreaction from $\chi$.
  However, the EoM for the perturbation $\delta\chi_{\bm{k}}$ exhibits instability modes due to the time-dependent frequency,
  \begin{equation}
    \omega_k^2\left(t\right)=k^2+\lambda^2a^2\left(t\right)\phi^2\left(t\right)=k^2+\lambda^2a^2\left(t\right)\Phi^2\left(t\right)\sin^2(m_{\phi}t).
  \end{equation}
We can rewrite Eq.~\eqref{eq:pertchi} as
\begin{equation}
  \label{eq:Mathieu}
\delta\chi_{\bm{k}}''+3\frac{H}{m_{\phi}}\delta\chi_{\bm{k}}'+\left[A\left(k\right)-2q\cos\left(2z\right)\right]\delta\chi_{\bm{k}}=0,
\end{equation}
where prime denotes differentiation with respect to $z=m_{\phi}t$,  $q=\lambda^2\Phi_i^2/4m_{\phi}^2$ is the parameter governing the resonance and $A\left(k\right)= k^2/m^2_{\phi}a^2+2q$. 
Neglecting Hubble friction term, Eq.~\eqref{eq:Mathieu} reduces to Mathieu equation with well-known solutions~\cite{Dolgov:1989us,Traschen:1990sw,Shtanov:1994ce,Kofman:1994rk,Kofman:1997yn}. 
An essential feature is existence~\cite{mclachlan1947theory} of growing instabilities $\delta \chi_k \propto e^{\mu_k z}$ with unstable modes within resonance band $\Delta k$. Here, $\mu_k$ is the exponential growth factor.
The presence of unstable solutions depends on the value of $A\left(k\right)$ and $q$. 
A broad resonance region exists for large $q$, wherein perturbation modes $\delta\chi_{\bm{k}}$ grow exponentially, causing rapid and non-uniform changes in the daughter field $\chi$.

In the case of instant preheating scenario the particle decay processes cannot be described by classical EoM. In order to effectively capture the effects of decays within classical EoM of fields, we introduce  the decay width $\Gamma$ as an additional friction term to EoM. Analogous strategy was also employed in some earlier works~\cite{Repond:2016sol, Fan:2021otj, Mansfield:2023sqp}.
The EoM of the inflaton field $\phi$ and the daughter field $\chi$ are then given by
\begin{align}
  \label{eq:infleom}
  \Ddot{\phi}+3H\Dot{\phi}-\frac{1}{a^2}\nabla^2\phi+\frac{\partial V}{\partial \phi}~&=0,\\
  \label{eq:chi2}
\Ddot{\chi}+\left(3H+\Gamma\right)\dot{\chi}-\frac{1}{a^2}\nabla^2\chi+\frac{\partial V}{\partial\chi}~&=0,
\end{align}
where $\Gamma=(\lambda y^2/8\pi)\left|\phi\right|$. In Ref.~\cite{Fan:2021otj,Mansfield:2023sqp} the authors analyze the model where the inflaton field $\phi$ and the daughter field $\chi$ are coupled through a trilinear term $\frac{1}{2}(M^2/f)\phi\chi^2$, which leads to tachyonic resonance. On the other hand, we analyze the model where the inflaton field $\phi$ and the daughter field $\chi$ are coupled through a quartic term $\frac{1}{2}\lambda^2\phi^2\chi^2$, leading to parametric resonance. From Eq.~\eqref{eq:infleom}, we see that the EoM of the inflaton field $\phi$ is not affected by the inclusion of the $\chi$'s decay. 
Compared with Eq.~\eqref{eq:chi1}, Eq.~\eqref{eq:chi2} has an additional friction term $\Gamma \dot{\chi}$ that delays the production of $\chi$ particle.
Approximating the fermionic field with a homogeneous radiation-like perfect fluid $\rho_{\text{R}}$, we obtain
\begin{equation}
  \label{eq:radiation} \dot{\rho}_{\text{R}}+3H\left(1+\omega\right)\rho_{\text{R}}=\left<\Gamma \dot{\chi}^2\right>,
\end{equation}
where $\left<...\right>$ denotes spatial average,  $\omega=p_{\text{R}}/\rho_{\text{R}}$ is the equation of state parameter of the radiation, $\omega=1/3$. 
The source of the radiaton energy density on the right hand side of Eq.~\eqref{eq:radiation} comes from the conservation of stress-energy tensor, $\nabla_{\mu}T^{\mu 0}=0$. 
The evolution of the scale factor $a$ is determined by,
\begin{align}
  \frac{\ddot{a}}{a}=&~-\frac{1}{6 M_{\text{Pl}}^2}(\bar{\rho}+3\bar{p}), \\
  \label{eq:EC}
  H^2=&~ \Big(\dfrac{\dot{a}}{a}\Big)^2 = \frac{\bar{\rho}}{3M_{\text{Pl}}^2},
\end{align} 
 where $\bar{\rho}$ denotes the mean energy density with
\begin{equation}
  \rho=K_{\phi}+K_{\chi}+G_{\phi}+G_{\chi}+V+\rho_{\text{R}},
\end{equation}
and $\bar{p}$ denotes the mean pressure with
\begin{equation}
  p=K_{\phi}+K_{\chi}-\frac{1}{3}\left(G_{\phi}+G_{\chi}\right)-V+\omega \rho_{\text{R}}.
\end{equation}
Here, $K_{\phi}=(1/2)\dot{\phi}^2$ is the kinetic energy 
 and $G_{\phi}=(1/2a^2)\Sigma_{i}\left(\partial_i\phi\right)^2$ is the gradient energy of $\phi$, with $K_\chi$ and $G_{\chi}$ denoting energies for $\chi$ analogously.
 
We have developed a software code to 
solve the EoMs of the classical fields by
conducting numerical lattice simulations using the finite-difference method and the fourth-order Runge-Kutta algorithm. 
To validate our code, we have simulated the preheating processes in $(1/2) m_{\phi}^2 \phi^2$ inflation in which 
the inflaton field $\phi$ couples the daughter field $\chi$ by the interaction term $\frac{1}{2}\lambda^2\phi^2\chi^2$. We have confirmed that there is good agreement between
dynamical evolution of fields and 
GW spectrum obtained from our simulations and the results of Ref.~\cite{Figueroa:2017vfa}.

\begin{figure}[t]
  \centering
  \begin{subfigure}[b]{0.48\textwidth}
\includegraphics[width=\textwidth]{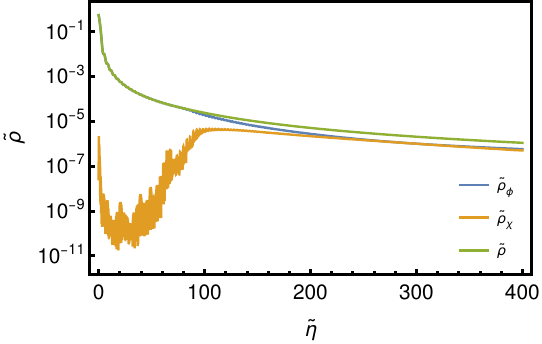}
    \label{fig:fig1subfig1}
  \end{subfigure}
  \begin{subfigure}[b]{0.48\textwidth}
\includegraphics[width=\textwidth]{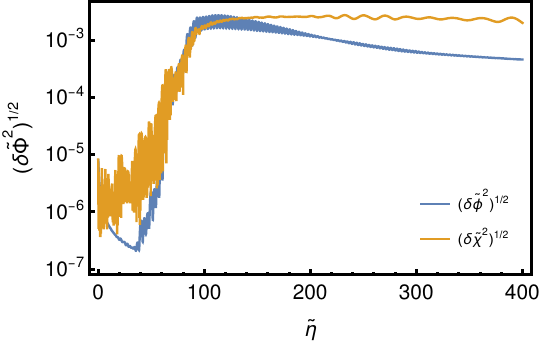}
    \label{fig:fig1subfig2}
  \end{subfigure}
  \vspace{-1em}
  \caption{Evolution of the total energy density $\tilde{\rho}$, 
  the energy density of the inflaton field $\tilde{\rho}_{\phi}$, 
  and the energy density of the daughter field $\tilde{\rho}_{\chi}$ as well as the root-mean-square $(\delta \tilde{\Phi}^2)^{1/2}$ of the inflaton field $\phi$ and the daughter field $\chi$ in the case of $y=0$.} 
  \label{fig:fig1}
\end{figure}

\subsubsection{Numerical lattice simulations}
In the following, we present the results of our simulations of preheating dynamics for instant preheating scenario. 
The mass of the inflaton field is set to $m_{\phi}=6\times 10^{-6} M_{\text{Pl}}$, 
a typical value in large field inflation models. As mentioned in Ref.~\cite{Figueroa:2017vfa}, 
if the resonance parameter $q\lesssim 6000$, when $10^{-4}<\lambda < 10^{-3}$, the system cannot be appropriately simulated.
On the other hand if $\lambda$ is too large, we cannot cover the proper ultraviolet (UV) dynamics of the system due to wide resonance band. Hence, we consider the coupling strength between the inflaton field $\phi$ 
and the daughter field $\chi$ to be $\lambda = 1.0 \times 10^{-3}$.

The initial conditions are chosen at the time $t_i$ when $H\left(t_i\right)=m_{\phi}$ holds exactly,
\begin{equation}
  \phi_i=2.32M_{\text{Pl}}, \quad \dot{\phi}_i=-0.78m_{\phi}M_{\text{Pl}}.
\end{equation}
To conduct numerical simulations on the lattice, we define new dimensionless variables
\begin{equation}
  \tilde{\phi}=\frac{\phi}{\phi_i},\quad \tilde{\chi}=\frac{\chi}{\phi_i},\quad \tilde{\eta}=m_{\phi}t,\quad \tilde{k}=\frac{k}{m_\phi}.
\end{equation}
The lattice size is set to $N = 256$, and the infrared momentum is $\tilde{k}_{\text{IR}}=5<q^{1/4}$, which ensures that the lattice can capture the relevant momentum modes in the simulations. We have simulated the system until $\tilde{\eta}=400$, 
since the system enters a stationary period at that point. We have confirmed that in all of our simulations the energy conservation of Eq.~\eqref{eq:EC} is satisfied at the level of $10^{-5}$. Hence, the approximation of fermion field as a uniform fluid has a minimal impact on the energy conservation.

In Fig.~\ref{fig:fig1}, we show the evolution of the total energy density $\tilde{\rho}$, 
the energy density of the inflaton field $\tilde{\rho}_{\phi}$, 
and the energy density of the daughter field $\tilde{\rho}_{\chi}$ as well as the root-mean-square root $(\delta \tilde{\Phi}^2)^{1/2}$ of the inflaton field $\phi$ and the daughter field $\chi$ when $y=0$. 
The field evolution can be divided into three phases. 
In the perturbation phase, when $\tilde{\eta}$ is between 0 to around $100$, the resonance perturbation $\chi_{\bm{k}}$ increases exponentially until the backreaction from the daughter field $\chi$ to the inflaton field $\phi$ becomes important.
 In the second stage, when $\tilde{\eta}$ is around $100$ to $200$, the backreaction becomes important and the evolution becomes highly non-linear. Subsequently, at the third stage, the system enters a stationary regime.

\begin{figure}[t]
  \centering
  \begin{subfigure}[b]{0.48\textwidth}
\includegraphics[width=\textwidth]{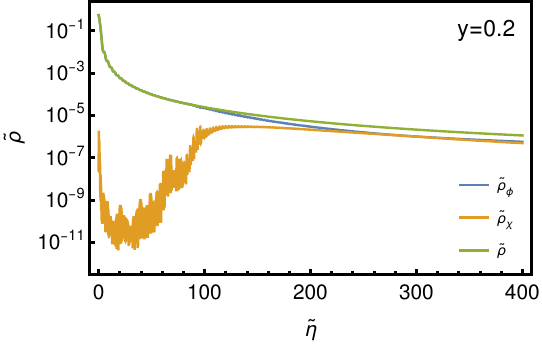}
  \end{subfigure}
  \begin{subfigure}[b]{0.48\textwidth}
\includegraphics[width=\textwidth]{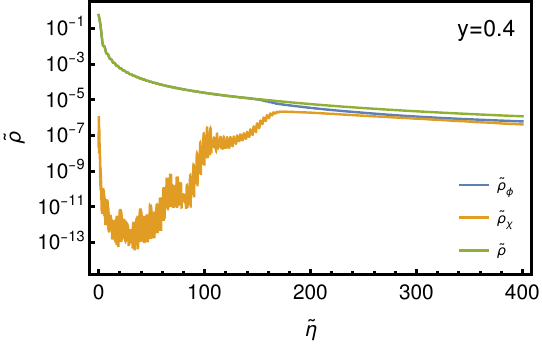}
  \end{subfigure}
  \begin{subfigure}[b]{0.48\textwidth}
\includegraphics[width=\textwidth]{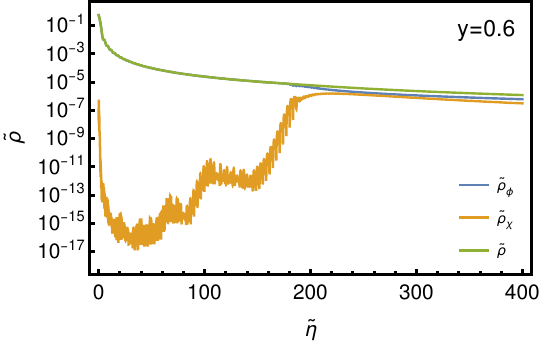}
  \end{subfigure}
  \begin{subfigure}[b]{0.48\textwidth}
\includegraphics[width=\textwidth]{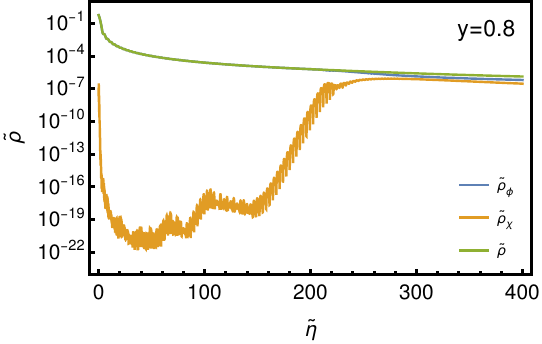}
  \end{subfigure}
  \caption{Evolution of the total energy density for various values of $y$.} 
  \label{fig:rho}
\end{figure}

Fig.~\ref{fig:rho} shows the evolution of the energy density when the value of the parameter $y \neq 0$. 
As the value of Yukawa coupling constant $y$ increases, the decay width $\Gamma$ also increases. 
Since $\Gamma$ term acts as a friction to the EoM of the daughter field $\chi$, the increase of the value of Yukawa coupling constant $y$ delays the growth of the energy density of the daughter field $\chi$, 
as can be observed.

\begin{figure}[t]
  \centering
  \begin{subfigure}[b]{0.48\textwidth}
\includegraphics[width=\textwidth]{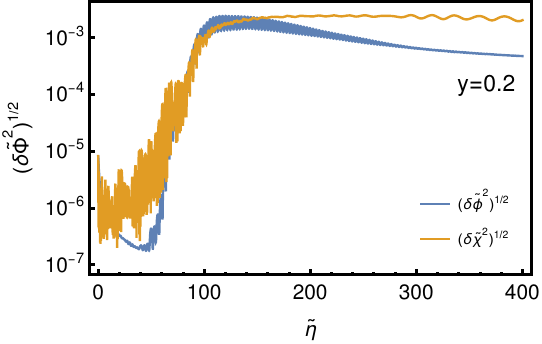}
  \end{subfigure}
  \begin{subfigure}[b]{0.48\textwidth}
\includegraphics[width=\textwidth]{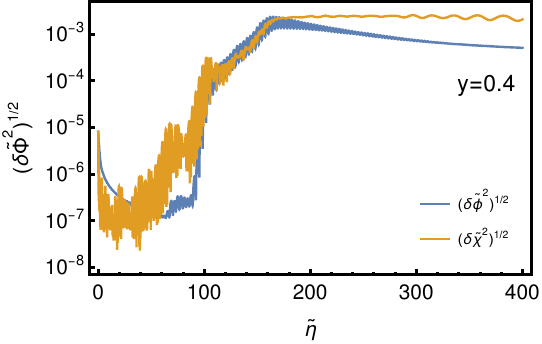}
  \end{subfigure}
  \begin{subfigure}[b]{0.48\textwidth}
\includegraphics[width=\textwidth]{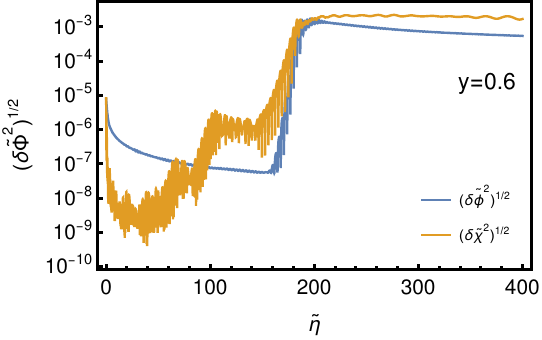}
  \end{subfigure}
  \begin{subfigure}[b]{0.48\textwidth}
\includegraphics[width=\textwidth]{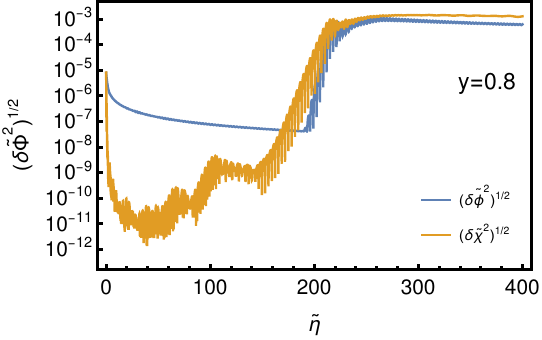}
  \end{subfigure}
  \caption{Evolution of the root-mean-square $(\delta \tilde{\Phi}^2)^{1/2}$ 
  of the inflaton field $\phi$ and the daughter field $\chi$ for various values of $y$.} 
  \label{fig:rms}
\end{figure}

In Fig.~\ref{fig:rms}, we display the evolution of the root-mean-square of the inflaton field $\phi$ and the daughter field $\chi$ for $y \neq 0$. 
Similar to the energy density of the $\chi$ field, the root-mean-square of the $\chi$ field also experiences a growth delay with increased Yukawa coupling constant $y$. Moreover, the time span for the growth of the root-mean-square of the $\chi$ field coincides with that of the energy density growth.

\begin{figure}[t]
  \centering   
  \begin{subfigure}[b]{0.48\textwidth}
\includegraphics[width=\textwidth]{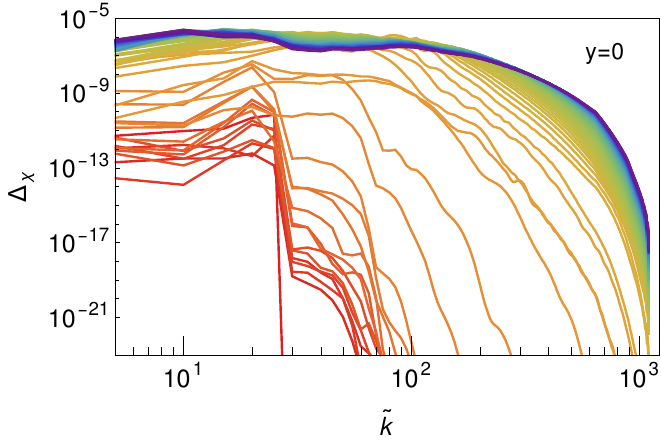}   
    \label{fig:p_chi_y0_0}    \vspace{-1em}
  \end{subfigure} 
  \begin{subfigure}[b]{0.48\textwidth}
\includegraphics[width=\textwidth]{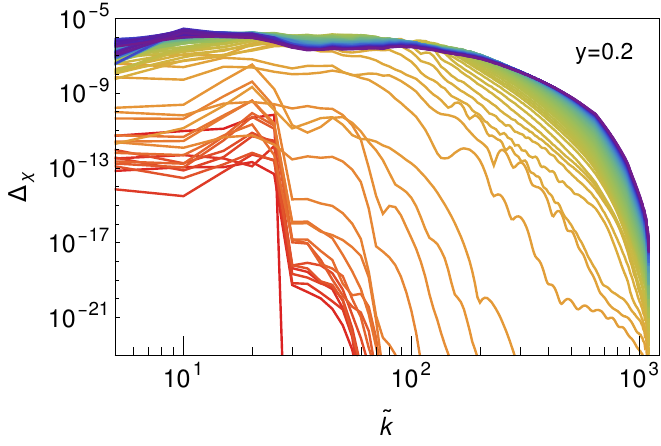}
  \end{subfigure}
  \begin{subfigure}[b]{0.48\textwidth}
\includegraphics[width=\textwidth]{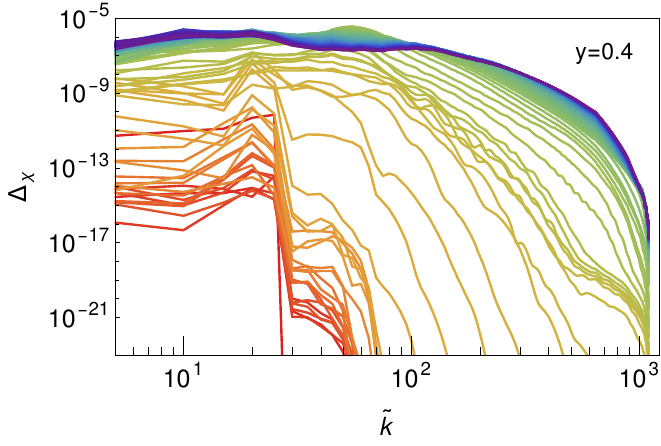}   
  \end{subfigure}
  \begin{subfigure}[b]{0.48\textwidth}   \includegraphics[width=\textwidth]{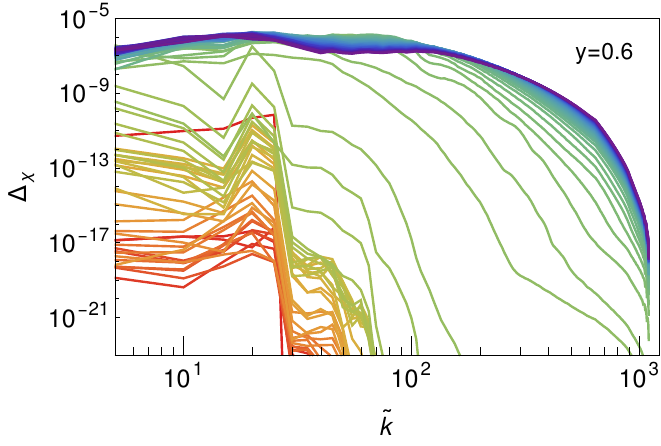}
  \end{subfigure}  
  \caption{Evolution of power spectrum of the daughter field $\chi$ for various values of $y$. 
  The spectra are plotted at time intervals $\tilde{\eta}=0,5,10,...$, 
  and evolved from the red curves (corresponding to earlier time) to the blue ones (corresponding to later times).} 
  \label{fig:p_chi}
\end{figure}

Fig.~\ref{fig:p_chi} shows the evolution of the power spectrum of the daughter field $\chi$. 
As shown in the left-top panel, at the onset of the simulation, in the time interval of $\tilde{\eta}$ from 0 to about 100, 
the momentum modes in the resonance bands of the power spectra of the $\chi$ field experience exponential growth. 
Then the system enters non-linear regime, in which the magnitude of peak stops increasing and the location of the peak moves towards UV region. Another characteristic behavior demonstrated in Fig.~\ref{fig:p_chi} is that as the value of Yukawa coupling constant $y$ increases, the growth of the power spectrum is delayed. Therefore, it takes more time for the power spectrum to grow for larger $y$.

\subsection{Production of gravitational waves}

Using results of our numerical simulations, combined with the EoM for the tensor perturbation, 
we can compute GW signals generated during the aforementioned dynamical processes. The EoM of tensor perturbations is given by
\begin{equation}
  \label{eq15}
\ddot{h}_{ij}+3H\dot{h}_{ij}-\frac{1}{a^2}\nabla^2 h_{ij}=\frac{2}{M^2_{\text{Pl}}}\Pi_{ij}^{TT},
\end{equation}
where $h_{ij}$ satisfies the transverse-traceless condition, $\partial_ih_{ij}=0$ and $\Sigma_i h_{ii}=0$. $\Pi_{ij}^{TT}$ is the transverse-traceless part of the anisotropic stress tensor, $\Pi_{ij}=\partial_i \phi \partial_j \phi+\partial_i \chi \partial_j \chi$. 
In the simulation, we solve the EoM of the variable $u_{ij}$ numerically,
\begin{equation}
  \ddot{u}_{ij}+3H\dot{u}_{ij}-\frac{1}{a^2}\nabla^2 u_{ij}=\frac{2}{M^2_{\text{Pl}}}\Pi_{ij},
\end{equation}
which is connected with the tensor perturbation $h_{ij}$ in the momentum space, $h_{ij}\left(\bm{k},t\right)=\Lambda_{ij,kl}(\hat{\bm{k}})u_{kl}\left(\bm{k},t\right)$.
The projection operator $\Lambda_{ij,lm}(\hat{\bm{k}})$ is defined as
\begin{equation}
  \Lambda_{ij,lm}(\hat{\bm{k}})=P_{il}(\hat{\bm{k}})P_{jm}(\hat{\bm{k}})-\frac{1}{2}P_{ij}(\hat{\bm{k}})P_{lm}(\hat{\bm{k}}), 
\end{equation}
where $P_{ij}=\delta_{ij}-\hat{\bm{k}}_i\hat{\bm{k}}_j$, and $\hat{\bm{k}}_i=\bm{k}_i/k$.  

The energy density power spectrum of GW is given by
\begin{equation}
\Omega_{\text{GW}}\left(k,t\right)=\frac{1}{\rho_c}\frac{d\rho_{\text{GW}}}{d \text{log}k}\left(k,t\right),
\end{equation}
where $\rho_c=3M_{\text{Pl}}^2H^2$ is the critical energy density of the Universe, and $\rho_{\text{GW}}$ is the energy density of the stochastic GW background, 
\begin{equation}
\rho_{\text{GW}}\equiv\frac{M_{\text{Pl}}^2}{4}\left<\dot{h}_{ij}\dot{h}_{ij}\right>,
\end{equation}
where $\left<\dots\right>$ denotes spatial average.

\begin{figure}[t]
  \centering
  \begin{subfigure}[b]{0.48\textwidth}    \includegraphics[width=\textwidth]{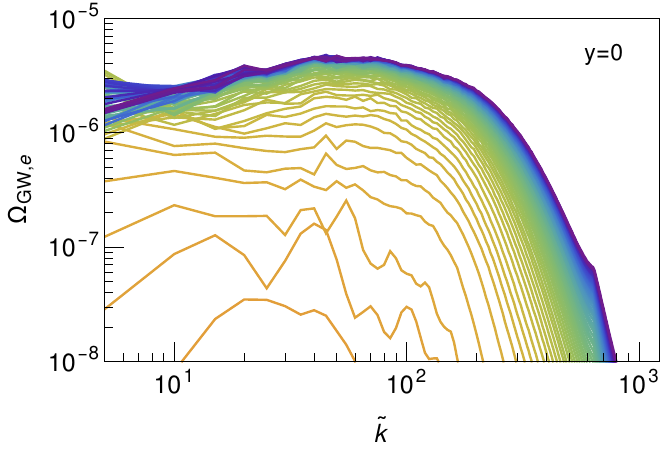}    \label{fig:GW_y0_0_2}\vspace{-1em}
  \end{subfigure}
  \hfill
  \begin{subfigure}[b]{0.48\textwidth}
\includegraphics[width=\textwidth]{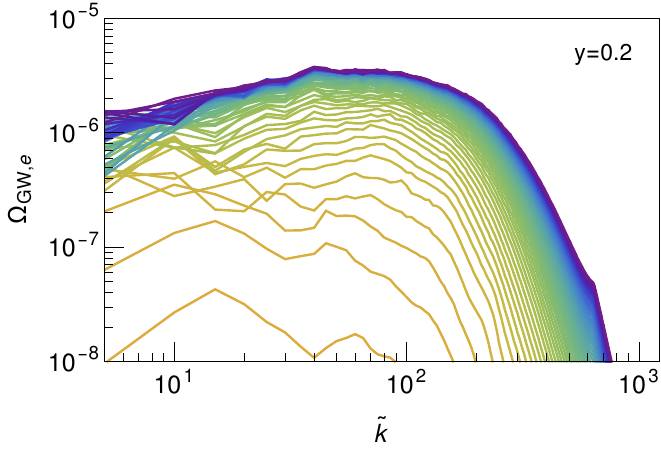}      \label{fig:GW_y0_2_2}\vspace{-1em}
  \end{subfigure}
  \vspace{0.5cm}
  \begin{subfigure}[b]{0.48\textwidth}
\includegraphics[width=\textwidth]{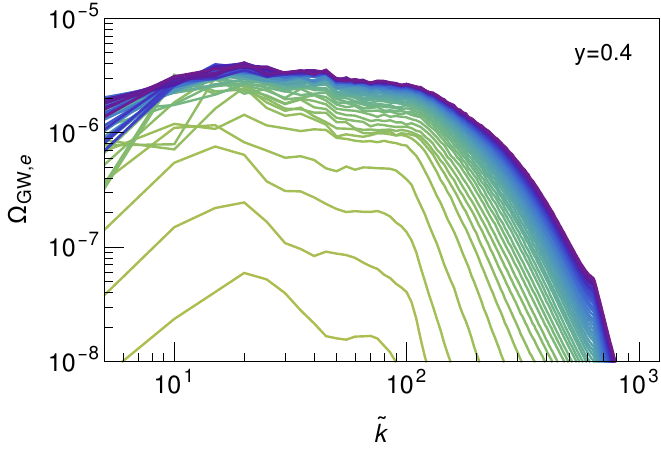}       \label{fig:GW_y0_4_2}\vspace{-1em}
  \end{subfigure}
  \hfill\vspace{-1em}
  \begin{subfigure}[b]{0.48\textwidth}
\includegraphics[width=\textwidth]{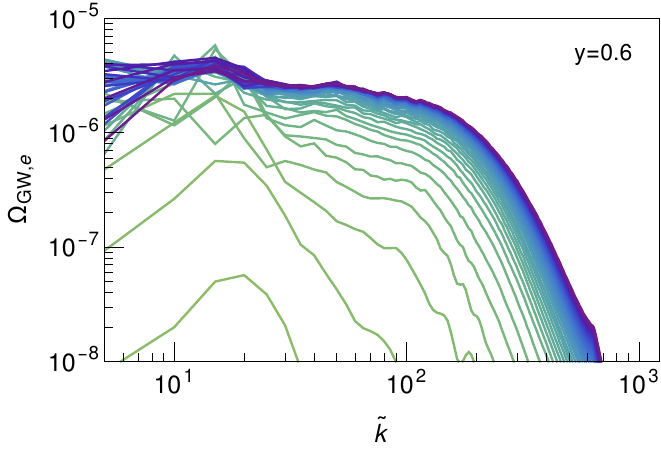}    
\label{fig:GW_y0_6_2}\vspace{-1em}
  \end{subfigure} \vspace{-1em}
  \caption{The evolution of GW energy spectra for different values of parameter $y$. 
  The spectra are plotted at time intervals $\tilde{\eta}=0, 5, 10, ...$, 
  and evolve from the red curves (corresponding to earlier times) to the blue ones (corresponding to later times).} 
  \label{fig:GW}
\end{figure}

In Fig.~\ref{fig:GW}, we present the evolution of GW energy spectra $\Omega_{\text{GW,e}}$ for different values of parameter $y$. 
We observe that as the parameter $y$ increases, the peak of the GW energy spectrum moves towards the infrared (IR) direction. 
When the value of the parameter $y$ is relatively large, for example, $y=0.9$, 
the peak position of GW is relatively small. If we capture the peak in the IR direction and 
ensure sufficient resolution in the UV region, 
larger grid size $N$ is required. Due to limitations of required computational resources, 
the cases with excessively large values of $y$ are beyond our analysis.

\begin{figure}[t]
    \centering
\includegraphics[width=0.48\textwidth,height=0.35\textwidth]{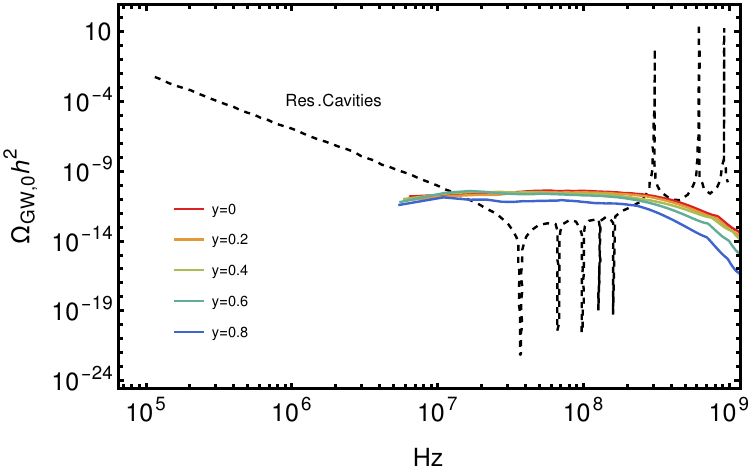}   \includegraphics[width=0.48\textwidth,height=0.35\textwidth]{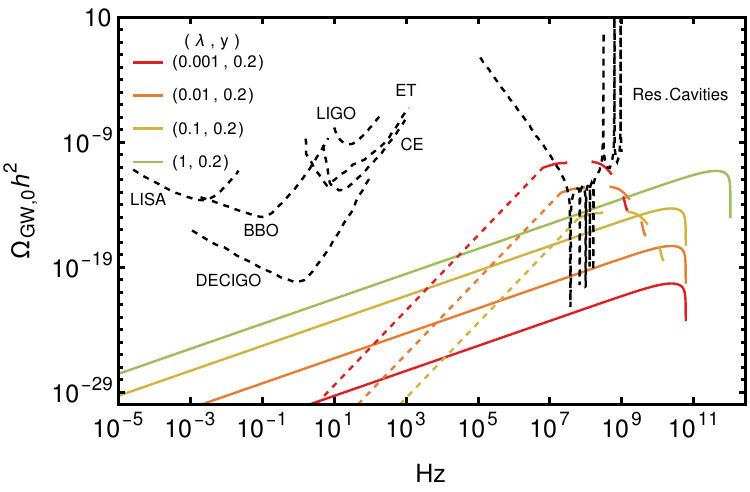}  
  \caption{[Left] GW energy spectra at present time from parametric resonance (solid colored lines) in scenario of instant preheating computed using numerical simulations for $\lambda=10^{-3}$ and various values of $y$. 
  [Right] GW energy spectra at present time from graviton bremsstrahlung emission (solid lines) and parametric resonance (dashed colored lines) for various values of $\left(\lambda,y\right)$. Sensitivity curves (dashed black lines) of proposed and existing GW detectors, including the fifth observing run (O5) of the aLIGO-Virgo detector network~\cite{LIGOScientific:2016fpe}, Laser Interferometer Space Antenna (LISA)~\cite{LISA:2017pwj}, Cosmic Explorer (CE)~\cite{Reitze:2019iox}, Einstein Telescope (ET)~\cite{Punturo:2010zz}, Big Bang Observer (BBO)~\cite{Harry:2006fi}, DECi-hertz Interferometer Gravitational wave Observatory (DECIGO)~\cite{Seto:2001qf} and resonance cavities~\cite{Herman:2020wao,Herman:2022fau} are also displayed.} 
  \label{fig:GW2today}
\end{figure}

In the left panel of Fig.~\ref{fig:GW2today}, we show GW energy spectra from parametric resonance at present time. 
This can be calculated as described in Ref.~\cite{Dufaux:2007pt}, 
\begin{align}
  f_{\rm GW}=&~\frac{k_0}{2\pi} = \frac{k/a_{\text{e}}}{2\pi}\frac{a_{\text{e}}}{a_0} \sim \frac{k/a_{\text{e}}}{\rho_{\text{e}}^{1/4}}\left(\frac{a_{\text{e}}}{a_{\rm R}}\right)^{1-\frac{3}{4}\left(1+\omega_{\rm R}\right)} \times (4\times10^{10})~\text{Hz}, \\ 
  \Omega_{\text{GW,0}}=&~\Omega_{\rm rad}\left(\frac{g_{*0}}{g_{*\rm R}}\right)^{1/3}\left(\frac{a_{\text{e}}}{a_{\rm R}}\right)^{1-3\omega_{\rm R}}\Omega_{\text{GW,e}},
\end{align}
where $a_{\text{e}}$ and $a_{\rm R}$ denote the scale factors $a$ at the end of simulation time $t=t_{\text{e}}$ and the beginning of radiation domination $t=t_{\rm R}$, respectively, and $\omega_{\rm R}$ is the equation of state parameter between $t_{\text{e}}$ and $t_{\rm R}$ while $g_*$ is the number of effective massless degrees of freedom. 
We will consider $g_{*\rm R}=100$ and $a_{\text{e}}=a_{\rm R}$ below. The GW energy spectra exhibit a broad peak in the frequency range of $10^7$~Hz to $10^{9}$~Hz, as shown in the left panel of Fig.~\ref{fig:GW2today}. Moreover, compared to the GW signals generated from particle decays through graviton bremsstrahlung, the frequency of these peaks is lower and may fall within the sensitivity range of proposed resonance cavities detectors~\cite{Herman:2022fau}, although their realization remains to be demonstrated.

Here, we introduce an analytical method for estimating the GW spectrum for larger values of $\lambda$. Let us consider the parameter region III in Fig.~\ref{fig:param} as discussed previously. The peak frequency and peak GW spectrum may be proportional to
\begin{align}
   & \frac{k_{\text{p}}}{a_\text{e}} \sim \sqrt{\lambda m_\phi \tilde\phi(t_{\rm P.R.})}\sim  \sqrt{\frac{\lambda \phi_i}{t_{\rm P.R.}}} \sim \sqrt{\frac{8\pi m_\phi^2}{y^2}},\\
   & f_{\text{p}} = \frac{k_{\text{p}}/a_{\text{e}}}{2\pi}\frac{a_{\text{e}}}{a_0} \sim \frac{1}{2\pi}\sqrt{\frac{\lambda\phi_i}{t_{\rm eq}}} \frac{a_{\rm eq}}{a_0}\sim \sqrt{\lambda}\sqrt{\frac{\phi_i}{M_{\rm Pl}}} \times 10^{11}\,{\rm Hz},\\
   & \Omega_{\rm GW,e}(k_{\rm p}) \sim \frac{\rho_\chi(t_{\rm P.R.})}{(k_{\rm p}/a_{\rm e})^2 M_{\rm Pl}^2}\sim \frac{\rho_\phi(t_{\rm P.R.})}{(k_{\rm p}/a_{\rm e})^2M_{\rm Pl}^2} \sim \frac{m_\phi^2}{\lambda^2y^2 M^2_{\rm Pl}},
\end{align}
where the subscript `eq' denotes the epoch of matter-radiation equality.
In the parameter region IV in Fig.~\ref{fig:param}, on the other hand, we have
\begin{align}
   & \frac{k_{\rm p}}{a_{\rm e}} \sim \sqrt{\lambda m_\phi \phi_i},\\
   & f_{\rm p} = \frac{k_{\rm p}/a_{\rm e}}{2\pi}\frac{a_{\rm e}}{a_0} \sim \frac{1}{2\pi}\sqrt{\frac{\lambda M_{\rm Pl}}{t_{\rm eq}}} \frac{a_{\rm eq}}{a_0} \sim \sqrt{\lambda} \times 10^{11}\,{\rm Hz},\\
   & \Omega_{\rm GW,e}(k_{\rm p}) \sim \frac{\rho_\chi(t_{\rm P.R.})}{(k_{\rm p}/a_{\rm e})^2 M_{\rm Pl}^2}\sim \frac{\rho_{\phi,i}}{(k_{\rm p}/a_{\rm e})^2M_{\rm Pl}^2} \sim \frac{m_\phi \phi_i}{\lambda M^2_{\rm Pl}}.
\end{align}
We first obtain the GW spectrum for $\lambda=10^{-3}$ and $y=0.2$ through numerical simulations. Then, using the relationship between the GW spectra and the parameters $\lambda$ and $y$, we scale the results as appropriate to estimate the spectra for larger values of $\lambda$. 

In the right panel of Fig.~\ref{fig:GW2today} we present the GW spectra at present time from instant preheating considering contributions from both the parametric resonance and the graviton bremsstrahlung processes associated with particle decays for various values of $\left(\lambda,y\right)$. We do not display the parametric resonance contributions for $\left(\lambda=1,y=0.2\right)$, because as we have demonstrated in Fig.~\ref{fig:param} no parametric resonance can occur for this case. Additionally, the low-frequency contributions displayed with colored dashed lines are derived based on the assumption that GW spectra are expected to be proportional to $\sim f^3$ at low frequencies~\cite{Liddle:1999hq,Easther:2006vd,Easther:2006gt,Dufaux:2007pt}. We observe that instant preheating scenario leads to two distinct GW signals produced in different frequency bands, originating from different sources.

\section{Conclusions and discussion} \label{sec:con}

Instant preheating frequently occurs in realistic models of inflation. We have identified two distinct types of GW signals characteristic of instant preheating scenario: one arising from parametric resonance and the other produced by graviton bremsstrahlung emission through particle decays. 
Detecting these two different bands of GW signals establishes unique possibilities for probing instant preheating. In particular, when the coupling constants take appropriate values, even Planck-scale particles can be produced. It predicts strongly enhanced bremsstrahlung GWs since its strength depends only on the mass of decaying particles.  
Therefore, detecting GW signals associated with gravitational bremsstrahlung emission from particle decays offers a means to probe Planck-scale physics, which is challenging to explore. 

Using numerical lattice simulations along with analytical estimates we have demonstrated the existence of two distinct GW signal peaks for a range of model coupling values: the peak from parametric resonance generically lies between $10^7$\,Hz and $10^9$\,Hz, while that from gravitational bremsstrahlung emission is typically between $10^{10}$\,Hz and $10^{11}$\,Hz. 
With larger inflaton and daughter particle coupling parameter $\lambda$, both the frequency and magnitude of the GW signals generated by gravitational bremsstrahlung increase, resulting in it being further distinguishable from the GWs generated from parametric resonance. We also found that in the preheating dynamics, as the Yukawa parameter $y$ that is also responsible for particle decays leading to graviton bremsstrahlung increases, the frequency of the GW generated by parametric resonance will also redshift.
Observing these dual GW signals provides a novel way to probe early Universe dynamics. 

However, we should note that we have treated the bremsstrahlung and parametric resonance GWs separately. In actual situation the distinction between them may be more vague. For example, bremsstrahlung GWs are produced even while parametric resonance is effective, which is difficult to calculate in lattice simulation. 
Thus more refined simulation or techniques may be required for obtaining more precise GW spectrum from instant preheating. Our results may be regarded as a first step to reach this goal.

\section*{Acknowledgment}

This work was supported by
World Premier International Research Center Initiative
(WPI), MEXT, Japan. 
W.Y.H was supported by the National Natural Science Foundation of China (NSFC) (No.
11975029, No. 12325503). 
This work was also supported by JSPS KAKENHI (Grant Numbers 24K07010 [K.N.] and 23K13109 [V.T.]).
Y.T. was supported by NSFC No. 12147103 and the Fundamental Research Funds for the Central Universities. This work was performed in part at the Aspen Center for Physics, which is supported
by the National Science Foundation grant PHY-2210452.

\bibliographystyle{utphys}
\bibliography{ref}

\end{document}